\documentclass[12pt]{article}

\ifx\pdfoutput\undefined
\usepackage[dvips,bookmarks]{hyperref}  
\else
\usepackage{hyperref}   
\fi

\usepackage{graphicx}
\usepackage{amsfonts}
\usepackage{amssymb,amsmath}

\hypersetup{colorlinks,bookmarksopen,bookmarksnumbered,citecolor=blue,
pdfstartview=FitH}
\def\hhref#1{\href{http://arxiv.org/abs/hep-th/#1}{hep-th/#1}} 
\def\mhref#1{\href{mailto:#1}{#1}}              

\def\bop#1{\setbox0=\hbox{$#1M$}\mkern1.5mu
\vbox{\hrule height0pt depth.04\ht0
\hbox{\vrule width.04\ht0 height.9\ht0 \kern.9\ht0
\vrule width.04\ht0}\hrule height.04\ht0}\mkern1.5mu}
\def\bo{{\mathpalette\bop{}}}                        
\def\frac#1#2{{\textstyle{#1\over#2}}}     

\parskip=\medskipamount         
\thicklines             
\begin{document}

\newcommand{\be}{\begin{equation}}
\newcommand{\ee}{\end{equation}}
\newcommand{\mx}{\mbox}
\newcommand{\mt}{\mathtt}
\newcommand{\p}{\partial}
\newcommand{\st}{\stackrel}
\newcommand{\al}{\alpha}
\newcommand{\bb}{\beta}
\newcommand{\ga}{\gamma}
\newcommand{\te}{\theta}
\newcommand{\de}{\delta}
\newcommand{\et}{\eta}
\newcommand{\ze}{\zeta}
\newcommand{\s}{\sigma}
\newcommand{\e}{\epsilon}
\newcommand{\om}{\omega}
\newcommand{\Om}{\Omega}
\newcommand{\la}{\lambda}
\newcommand{\La}{\Lambda}
\newcommand{\ti}{\widetilde}
\newcommand{\ih}{\hat{i}}
\newcommand{\jh}{\hat{j}}
\newcommand{\kh}{\widehat{k}}
\newcommand{\lh}{\widehat{l}}
\newcommand{\eh}{\widehat{e}}
\newcommand{\ph}{\widehat{p}}
\newcommand{\qh}{\widehat{q}}
\newcommand{\mh}{\widehat{m}}
\newcommand{\nh}{\widehat{n}}
\newcommand{\Dh}{\widehat{D}}
\newcommand{\2}{{\textstyle{1\over 2}}}
\newcommand{\3}{{\textstyle{1\over 3}}}
\newcommand{\4}{{\textstyle{1\over 4}}}
\newcommand{\8}{{\textstyle{1\over 8}}}
\newcommand{\6}{{\textstyle{1\over 16}}}
\newcommand{\ra}{\rightarrow}
\newcommand{\lra}{\longrightarrow}
\newcommand{\Ra}{\Rightarrow}
\newcommand{\im}{\Longleftrightarrow}
\newcommand{\hs}{\hspace{5mm}}
\newcommand{\bea}{\begin{eqnarray}}
\newcommand{\eea}{\end{eqnarray}}
\newcommand{\NP}{{\em Nucl.\ Phys.\ }}
\newcommand{\AP}{{\em Ann.\ Phys.\ }}
\newcommand{\PL}{{\em Phys.\ Lett.\ }}
\newcommand{\PR}{{\em Phys.\ Rev.\ }}
\newcommand{\PRL}{{\em Phys.\ Rev.\ Lett.\ }}
\newcommand{\PRP}{{\em Phys.\ Rep.\ }}
\newcommand{\CMP}{{\em Comm.\ Math.\ Phys.\ }}
\newcommand{\MPL}{{\em Mod.\ Phys.\ Lett.\ }}
\newcommand{\IJMP}{{\em Int.\ J.\ Mod.\ Phys.\ }}

\begin{titlepage}
\setcounter{page}{0} 
\begin{flushright}
YITP-SB-06-54\\ Nov. 28, 2006\\
\end{flushright}

\begin{center}
{\centering {\LARGE\bf Yang-Mills Gauge Conditions from Witten's Open String Field Theory \par} }

\vskip 1cm {\bf Haidong Feng\footnote{E-mail address:
\mhref{hfeng@ic.sunysb.edu}}, Warren Siegel\footnote{E-mail
address: \mhref{siegel@insti.physics.sunysb.edu}}} \\ \vskip 0.5cm

{\it C.N. Yang Institute for Theoretical Physics,\\
State University of New York, Stony Brook, 11790-3840 \\}

\end{center}

\begin{abstract}
We construct the Zinn-Justin-Batalin-Vilkovisky action for tachyons and gauge bosons
from Witten's 3-string vertex of the bosonic open string without
gauge fixing. Through canonical transformations, we find the off-shell,
local, gauge-covariant
action up to 3-point terms, satisfying the usual field theory gauge transformations. Perturbatively, it
can be extended to higher-point terms.
It also gives a new gauge condition in field
theory which corresponds
to the Feynman-Siegel gauge on the world-sheet.
\end{abstract}

\end{titlepage}

\section{Introduction}
In the usual (super)string theory, the external states for gauge bosons
are introduced by vertex operators with a gauge condition $b_0 = 0$ to have the
right conformal weight. Although this constraint can be relaxed to find gauge-covariant
unintegrated vertex operators, we still need the gauge-invariant equation of motion
for the free vectors \cite{fs1} and the effective action is valid only on-shell \cite{fs2}.
On the other hand, string field theory (SFT), the second-quantized approach to string theory,
can be used for an off-shell analysis. A complete description of
interacting strings and string fields was presented in the light-cone gauge
\cite{Goddard} and generalized to the super case \cite{Green}.
A covariant, gauge-invariant formulation of the bosonic open string field theory was given by Witten \cite{witten}, based on the relation found between gauge transformations of the fields and first-quantized Becchi-Rouet-Stora-Tyutin transformations in the free action \cite{Siegel}.  It was made more
concrete by several groups: The explicit operator construction of the string field
interaction was presented \cite{Gross-Jevicki-123}; string field theory geometry was
formulated by writing each term in the action as an expectation value in the 2D conformal
field theory on the world surface \cite{lpp}; the tensor constructions were analyzed
from first principles \cite{Gaberdiel-Zwiebach}; etc.

To calculate with the action for Witten's open string field theory,
it is helpful to fix the gauge. A particularly useful gauge
choice is the Feynman-Siegel gauge
\be b_0 | \Psi \rangle =0 \ee
The antifields in the string field expansion, which are associated with states
that have a ghost zero-mode $c_0$, are taken to vanish. Then the
action from the viewpoint of quantum field theory is gauge fixed, while it is not
clear what kind of gauge condition is applied. So we can only guess the  action for these
states (for example, the origin of the $\phi A^2$ term is not clear for
the lack of gauge covariance) but are not able to write it
down gauge invariantly. The simplest way to accomplish this is to find the Zinn-Justin-Batalin-Vilkovisky action \cite{ZJ,BV} with all antifields. The ZJBV
formalism was first
developed to deal with the renormalization of gauge theories, but follows naturally from any field theory action whose kinetic operator is expressed as the first-quantized BRST operator \cite{Siegel1}. It allows the
handling of very general gauge theories, including those with open or reducible
symmetry algebras. The ZJBV action includes both the usual gauge-invariant action and the definition of the gauge (BRST) transformations. Here we will start from this ZJBV action for SFT and,
through some canonical transformations (including field redefinitions and gauge transformations of both fields and antifields), get
the explicit gauge-covariant action (and gauge transformations) for tachyons and massless vectors up to 3-point terms.
We will show for the first
time that it is just usual Yang-Mills coupled to scalars, plus
$F^3$ and $\phi F^2$ interactions. These specific canonical transformations will tell us the gauge condition on the fields
corresponding to Feynman-Siegel gauge on the world-sheet.
Another advantage of this mechanism
is that we pushed all nonlocal factors in 3-point interactions
to higher-point interactions and make the 3-point interactions
just the usual local YM form. But, as a price, there will be all possible higher-point
interactions (nonrenormalizable in ordinary field theory), as shown in section \ref{5}.

The outline of this letter will be as follows: In section \ref{2}, we will briefly review
Witten's open string field theory; in section \ref{3}, we will give an introduction to the  ZJBV formalism
for Yang-Mills theory; in section \ref{4}, we will calculate the full ZJBV  action for tachyons and massless vectors from
Witten's open string field theory without the Feynman-Siegel gauge, and
find the suitable canonical
transformations to get back the ``original" gauge-invariant action to lowest order
in the Regge slope for the nonlocal exponential factors; in
section \ref{5}, we will perform further transformations to absorb the nonlocal
factors in 3-point interactions and push them to higher-point interactions, which
will give the usual local action up to 3-point terms; \hyperlink{end}{finally}, we will give
some discussion and conclusions.

\section{Witten's 3-string vertex}\label{2}

In string field theory, the 3-string interaction can be interpreted as
\be \langle h_1 [\vartheta_A] h_2 [\vartheta_b] h_3 [\vartheta_c] \rangle =
\langle V_{123} (| A \rangle_1 \otimes
|B \rangle_2 \otimes | C \rangle_3 ) \ee
where $\vartheta_i$ is the vertex operator for each external state and
$h_i(z)$ is the conformal mapping from each string state to the complex plane.
In Witten's bosonic open string field theory, strings interact by identifying the right half
of each string with the left half of the next one.
The conformal mapping for this interactive world-sheet geometry
can be expressed as
\be h_1 (z) = e^{i\frac{2\pi}{3}} h(z), \quad h_2 = h(z), \quad
h_3 = e^{-i\frac{2\pi}{3}} h(z) \ee
where
\be h(z) = \big( \frac{1-iz}{1+iz} \big)^{\frac{2}{3}} \ee
Then the action is
\begin{equation}
S = \langle V_2 | \Psi, Q \Psi \rangle
  + \frac{g}{3} \langle V_3 | \Psi, \Psi, \Psi \rangle
\end{equation}
where $Q$ is the usual string theory BRST operator. Using
the string oscillation modes $\alpha_n$ of the matter sector and
$b_n, c_n$ of the ghost sector, the two-string ``vertex" is
\begin{eqnarray}\label{v2}
\langle V_2 |  & = & \delta^D (p_1 + p_2)
\left( \langle 0; p_1 | \otimes \langle 0; p_2 | \right)
(c^{(1)}_0 + c^{(2)}_0) \nonumber \\
&  & \quad \times \exp \left(
\sum_{n = 1}^{ \infty} (-1)^{n+1}
[\alpha^{(1)}_n \alpha^{(2)}_n+c^{(1)}_n b^{(2)}_n+c^{(2)}_n b^{(1)}_n] \right)\,
\end{eqnarray}
and the 3-string vertex associated with the three-string
overlap can be written as
\begin{eqnarray}\label{V3witten}
\langle V_3| &=& {\mathcal N}  \delta^D (p_1 + p_2 + p_3)
(\langle 0| c_{-1} c_0 )^{(3)} )
(\langle 0| c_{-1} c_0 )^{(2)})  (\langle 0| c_{-1} c_0 )^{(1)}) \nonumber \\
&& \times \exp \Bigl( \sum_{r,s=1}^3 \sum_{n,m\geq 1} \,
{1\over 2} \alpha_m^{(r)} N^{rs}_{mn} \alpha_n^{(s)}
+ p^{(r)} N^{rs}_{0m} \alpha_m^{(s)} + {1\over 2} N_{00} \sum_{r=1}^3 (p^{(r)})^2 \Bigr)
\nonumber \\ && \times \exp \Bigl( \sum_{r,s=1}^3 \sum_{m\geq 0\atop n\geq 1}
b_m^{(r)} X^{rs}_{mn} n c_n^{(s)} \Bigr)
\end{eqnarray}
with the normalization factor ${\mathcal N}= 3^{9/2}/2^6$ \cite{Taylor}. Because we will focus
on the fields and antifields up to oscillation modes 1, the only relevant Neumann
coefficients are
\bea\label{Neumann}
& N^{11}_{11} = N^{22}_{11} = N^{33}_{11} = - \frac{5}{27} \nonumber \\
& N^{12}_{11} = N^{23}_{11} = N^{31}_{11} = \frac{16}{27} \nonumber \\
& N^{12}_{01} = - N^{13}_{01} = N^{23}_{01}
= - N^{21}_{01}= N^{31}_{01} = - N^{32}_{01} = -\frac{2 \sqrt{3}}{9} \nonumber \\
& N^{11}_{00} = N^{22}_{00} = N^{33}_{00} = -\frac{1}{2} \ln (27/16) \nonumber \\
& N^{11}_{01} = N^{22}_{01} = N^{33}_{01} =0
\eea
for the matter sector and
\bea
& X^{11}_{11} = X^{22}_{11} = X^{33}_{11} = -\frac{11}{27} \nonumber \\
& X^{12}_{11} = X^{23}_{11} = X^{31}_{11} = X^{21}_{11} = X^{32}_{11} = X^{13}_{11} = -\frac{8}{27} \nonumber \\
& X^{12}_{01} = - X^{13}_{01} = X^{23}_{01}
= - X^{21}_{01}= X^{31}_{01} = - X^{32}_{01} = -\frac{4 \sqrt{3}}{9} \nonumber \\
& X^{11}_{01} = X^{22}_{01} = X^{33}_{01} =0
\eea
for the ghost sector.

Usually, the three-string interactions are calculated in the Feynman-Siegel gauge
\be b_0 |\Psi \rangle = 0 \ee
Then what we get is the gauge-fixed  action, and
the gauge condition for this action was never clear. Also we will get some $\phi A^2$
interactions whose origin was not obvious due to the
lack of gauge covariance. In the next section,
we will construct the ZJBV action from string field
theory to study the gauge condition from the aspect of field theory.

\section{ZJBV}\label{3}
In the usual Hamiltonian formalism for a phase space ($q$, $p$),
the Poisson bracket, which is useful for studying symmetry properties and
relates to the commutator of the quantum theory,
can be defined. In gauge field theory, there is a similar interpretation where the
fields (including ghosts) correspond to $q$ and the antifields (with opposite statistics) to $p$.
In the YM case (including scalars), $\phi, A_{\mu}, C , \widetilde{C}$ are fields and $\phi^*,
A^*_{\mu}, C^* , \widetilde{C}^*$ are antifields. As a
generalization of the Poisson bracket, the ``antibracket" $(f(\Phi) , g(\Phi) ) = f \circ g$
is introduced \cite{ZJ}:
\be \circ = \int dx (-1)^I \left( {\overleftarrow{\delta} \over \delta \phi^*_I}
{\delta \over \phi^I} +  {\overleftarrow{\delta} \over \delta \phi^I}
{\delta \over \phi^*_I} \right) \ee
It has the following useful properties:
\bea\label{property}
& (f,ga) = (f,g)a, \quad (af,g) = a(f,g) \nonumber \\
& (f,g) = -(-1)^{(f+1)(g+1)}(g,f) \nonumber \\
& (f,gh) = (f,g)h + (-1)^{(f+1)g}g(f,h) \nonumber \\
& (-1)^{(f+1)(h+1)}(f,(g,h)) + cyc. = 0
\eea
The existence of a bracket with these properties allows the
definition of a Lie derivative, ${\mathcal L}_A B \equiv (A,B) $ and
a unitary transformation
\be S' = e^{{\mathcal L_G}} S = S + {\mathcal L_G} S + \frac{1}{2!}
{\mathcal L_G} {\mathcal L_G} S + \cdots \ee
For the example we are going to discuss, the antibrackets for fields and antifields
are:
\be (A^*_{\mu}, A_{\nu} ) = \eta_{\mu \nu}, \quad (\phi^*, \phi) = 1, \quad
(C, C^*) = 1, \quad (\widetilde{C} , \widetilde{C}^*) = 1 \ee

The general Lagrangian path integral for BRST
quantization is
\be {\mathcal A} = \int D \psi^I e^{-iS_{gf}}, \quad S_{gf} = e^{{\mathcal L}_{\Lambda}} S_{ZJBV}| \ee
where $S_{gf}$ is evaluated at all antifields $\psi^* =0$. Expanding the ZJBV action in antifields,
using $\psi_m$ and $\psi_{nm}$ to indicate all minimal and non-minimal
fields,
\be S_{ZJBV} = S_{gi} + (Q \psi_m) \psi_m^* + \psi_{nm}^* \psi_{nm}^* ,\ee
then
\be S_{gf} = e^{{\mathcal L}_{\Lambda}} S_{ZJBV}| =
S_{gi} + (\delta \Lambda / \delta \psi_{m}) (Q \psi_m) +
(\delta \Lambda / \delta \psi_{nm})^2 \ee
where $S_{gi}$ and $\Lambda$ depend only on coordinates $\psi^I$.
Also, the BRST transformations can be written as
$\delta_Q \psi^I = (S_{ZJBV} , \psi^I )$. Gauge independence requires
\be (-1)^I \frac{\delta^2 S_{ZJBV}}{\delta \psi^*_I \delta \psi^I} + i \frac{1}{2}
(S_{ZJBV}, S_{ZJBV}) = 0 \ee
which is called the ``quantum master equation". It is the approach
to BRST of Zinn-Justin, Batalin, and Vilkovisky (ZJBV).

To see the equivalence of the ZJBV combination of the gauge-invariant
action with the BRST operator to ordinary BRST,
here is an example, pure Yang-Mills theory. The ZJBV action in YM can be written as
$$ S_{ZJBV} = - F_{\mu \nu} F^{\mu \nu} - 2 (\widetilde{C}^*)^2
- 2i [\triangledown_{\mu},C] A^{*\mu} + C^2 C^* $$
We have the usual BRST transformations of fields from $Q\psi = (S, \psi)$:
\be QA_{\mu} = - 2i [\triangledown_{\mu},C], \quad QC = -C^2,
\quad Q\widetilde{C} = 4 \widetilde{C}^*, \quad Q\widetilde{C}^* = 0 \ee
Taking
\be \Lambda = tr \int \frac{1}{4} \widetilde{C} f(A) , \ee
we find the usual gauge fixed action
\be  S_{gf} = S_{gi} - \frac{1}{4} f(A)^2 - \frac{i}{2} \widetilde{C}
\frac{\partial f}{\partial A} \cdot [\triangledown , C] \ee
as from the usual BRST formalism.

\section{The gauge covariant action}\label{4}

In this section, we will use Witten's 3-string vertex to get the interactions
for tachyons and vectors without the Feynman-Siegel gauge. The  action
will be in the ZJBV formalism including fields and antifields.
From this ZJBV action, through some canonical transformations, we can get the gauge invariant action back. Observing the
forms of these transformations, we will be able to tell which gauge condition
in field theory corresponds to the Feynman-Siegel gauge in Witten's
string field theory.

In string field theory, the general external state (without $b_0 =0$) is
\be |\psi \rangle  = (C + \phi c_1 + A \cdot a_{-1} c_1 + \widetilde{C} c_{-1} c_1 +
\widetilde{C}^* c_0 + \phi^* c_0 c_1 + A^* \cdot a_{-1} c_0 c_1 +
C^* c_{-1} c_0 c_1 + \cdots ) | 0,k \rangle \ee
It gives the free terms and 3-point interactions for
tachyons, YM gauge bosons, ghosts, antighosts, and their antifields.
The free part is
\bea\label{S_2} S^{ZJBV}_2 &=& \langle V_2 | \Psi, Q \Psi \rangle \nonumber \\
 & = & - \frac{1}{2} \phi (\bo + 2 ) \phi
- \frac{1}{2} A_{\mu} \bo A^{\mu} + \widetilde{C} \bo C - 2i (\partial_{\mu} C) A^{*\mu}
\nonumber \\ && -2 (\widetilde{C}^*)^2 - 2i (\partial \cdot A) \widetilde{C}^*
\eea
and the interaction part is
\bea\label{S_3} S^{ZJBV}_3 = \frac{g}{3} \langle V_3 | \Psi, \Psi, \Psi \rangle = S_3^{(0)} +
S_3^{(1)} + S_3^{(2)} + S_3^{(3)}
\eea
where (to lowest order in Regge slope for those nonlocal factors $e^{\frac{1}{2} N_{00}^{rr}
(P^2_i +  m_i^2)}$; we will discuss them in the next section)
\bea S_3^{(0)} &=& \frac{1}{3} \phi^3 + \phi A^2 + (\widetilde{C}^*)^2 \phi
-\frac{1}{2} [\widetilde{C}, C] \phi + \frac{1}{2} \{ \widetilde{C}, C \} \widetilde{C}^*
\nonumber \\
&& + \{C, \phi^*\} \phi + C^2 C^* + [\phi^* , C] \widetilde{C}^* + [A_{\mu}, C] A^{*\mu} \\
S_3^{(1)} &=& \frac{i}{2} \partial_{\mu} \phi [A^{\mu}, \phi] + \frac{i}{4} (\partial_{\mu} A_{\nu} - \partial_{\nu} A_{\mu}) [A^{\mu}, A^{\nu}] \nonumber \\
&& + \frac{i}{2} [\widetilde{C}^* , \partial_{\mu} \widetilde{C}^*] A^{\mu} +
\frac{i}{4} \widetilde{C} [A_{\mu} , \partial^{\mu} C] -
\frac{i}{4} \partial^{\mu} \widetilde{C} [A_{\mu} , C]
\nonumber \\
&& + \frac{i}{2} \phi^* ( \{ \partial_{\mu} C , A^{\mu} \} +
\partial_{\mu} \{ C , A^{\mu} \} ) \nonumber \\
&& + \frac{i}{2} A^*_{\mu} ( [C, \partial^{\mu} \widetilde{C}^*]
- [\partial^{\mu} C, \widetilde{C}^*] ) \nonumber \\
&& + \frac{i}{2} A^*_{\mu} ( \{ C, \partial^{\mu} \phi]
- [\partial^{\mu} C, \phi] ) \\
S_3^{(2)} &=& \phi (\partial_{\mu} A_{\nu}) (\partial^{\nu} A^{\mu})
+ \frac{1}{2} \phi \{ \partial_{\mu} (\partial \cdot A), A^{\mu} \}
+ \frac{1}{4} \phi (\partial \cdot A)^2 \nonumber \\
&& + ( \frac{1}{2} [\partial_{\nu} C, \partial_{\mu} A^{\nu}]
- \frac{1}{2} [\partial_{\mu} \partial_{\nu} C, A^{\nu}] \nonumber \\
&& \quad + \frac{1}{4} [C, \partial_{\mu}(\partial \cdot A)]
- \frac{1}{4} [\partial_{\mu} C, (\partial \cdot A) ])A^{*\mu} \\
S_3^{(3)} &=& \frac{i}{6} (\partial^{\mu} \partial^{\nu} A_{\lambda})
[A_{\mu}, \partial^{\lambda} A_{\nu}]
+ \frac{i}{24} \partial^{\mu} (\partial \cdot A) [\partial^{\nu} A_{\mu} , A_{\nu}]
\eea

This gives the gauge fixed action after setting antifields to zero. Before setting them to zero,
it is related to the usual ZJBV action by a canonical (with respect to the antibracket) transformation.
Since such transformations can mix fields and antifields, the transformation itself
(followed by setting antifields to zero) is one way to define the gauge-fixing procedure
in this formalism.  So, one way to find the gauge invariant action is to undo this transformation.

Another way is to take this action with antifields, drop all fields with nonvanishing ghost number,
and then eliminate the remaining zero-ghost-number antifields (Nakanishi-Lautrup fields) by their
equations of motion.  However, the resulting action is kind of messy and has unusual gauge transformations.

The advantage of working with the entire ZJBV action is that it contains both the gauge invariant action and the gauge (BRST) transformations. Furthermore, canonical transformations perform field redefinitions (including antifield redefinitions that define the gauge fixing) in a way that preserves the (anti)bracket (as in ordinary quantum mechanics).  Thus, we look for canonical transformations that produce the standard form for gauge transformations of the fields, a well as eliminate terms in the action that could normally be ignored ``on shell".

Notice there are antifield-independent terms from gauge fixing in
the ZJBV action of (\ref{S_2}) and (\ref{S_3}).
So we have to find transformations to  ``undo" the gauge fixing. For example, the gauge
transformation generated by $-\frac{i}{2} (\partial \cdot A) \widetilde{C}$ will
cancel the gauge fixing term $\widetilde{C} \bo C$ because
$(-\frac{i}{2} (\partial \cdot A) \widetilde{C}, - 2i (\partial_{\mu} C) A^{*\mu})
= -\widetilde{C} \bo C$. Also notice that the ZJBV actions of (\ref{S_2}) and (\ref{S_3})
don't give the usual gauge transformations (from terms linear
in antifields), so we also look for transformations to
give them the usual form. For instance, the term $[\phi^* , C] \widetilde{C}^*$ will give
unusual contributions for gauge transformations of $\phi$ and $\widetilde{C}$, but it
can be canceled through the field redefinition generated by $\frac{1}{4} [\phi^*, C] \widetilde{C}$.
We also look for terms that generate field redefinitions that
cancel cubic antifield-independent terms that are proportional to the
linearized field equations. For example, $\frac{1}{2} A^2 \phi^*$ will generate 
the counter term $-\phi A^2 - \frac{1}{2} (\bo \phi) A^2$, which converts $\phi A^2$
into $- \frac{1}{2} (\bo \phi) A^2$, which will be part of the covariant interaction 
$\phi F_{\mu \nu} F^{\mu \nu}$.

The calculation is straightforward, but to find the complete transformation
we need more steps, because some transformations applied to cancel terms
we don't want will have byproducts to be canceled by further transformations. 
The complete transformation is given as follows:
First, make the transformation generated by
\be G_{g} = -\frac{i}{2} (\partial \cdot A) \widetilde{C}
+ \frac{1}{16} \widetilde{C} [ \bo A_{\mu} + \partial_{\mu} (\partial \cdot A) , A^{\mu}] +
\frac{1}{8} \widetilde{C}^2 C + \frac{1}{16} (\partial_{\mu} \widetilde{C})^2 C
- \frac{i}{8} \widetilde{C} \{ \partial \cdot A, \phi \}
\ee
to  ``undo" the gauge fixing. It is independent of antifields, and so can be
identified with gauge fixing. Then we make the transformation
\bea G_0 &=& \frac{1}{4} [\phi^*, C] \widetilde{C} + \frac{1}{2} A^2 \phi^*
+ \frac{1}{4} \{C, \phi \} C^* +
\frac{1}{8} \{ \phi, \widetilde{C}^* \} \widetilde{C}
- \frac{1}{2} \{ \phi , A^*_{\mu} \} A^{\mu} \nonumber \\
&& - \frac{i}{4} A^*_{\nu} [\partial^{\nu} A^{\mu}, A_{\mu}]
+ \frac{i}{8} \{ C, A^*_{\mu} \} (\partial^{\mu} \widetilde{C})
-\frac{i}{8} [ \widetilde{C}^* , A_{\mu} ] (\partial^{\mu} \widetilde{C})
\eea
This generator is linear in antifields, and so can be identified with a field redefinition.
(However, there is some subtlety in that the Nakanishi-Lautrup fields in this form of ZJBV appear as antifields $\widetilde{C}^*$.)
As the result of the above transformations, the action (up to 3-point terms and lowest order in
Regge slope) can be written as
\bea\label{S_gi} S &=& S_2 + S_3 \nonumber \\
&=& \frac{1}{2} [\nabla_{\mu} , \phi] [\nabla^{\mu} , \phi] - \phi^2
- F_{\mu \nu} F^{\mu \nu} - 2i [\triangledown_{\mu}, C] A^{*\mu}
-2 (\widetilde{C}^*)^2 + \{ C, \phi^* \} \phi + C^2 C^* \nonumber \\
 && + \frac{1}{3} \phi^3 + 2 \phi F_{\mu \nu} F^{\mu \nu}
- \frac{4}{3} F_{\mu}^{\nu} F_{\nu}^{\lambda} F_{\lambda}^{\mu}
\eea
with $\triangledown_{\mu} = \partial_{\mu} + \frac{i}{2} A_{\mu}$.
Now it is explicitly gauge covariant (to this order) even off-shell! Thus
the $F^3$ interaction appears explicitly (which was done only on shell before),
and a new gauge invariant interaction term $\phi F^2$ is found. Furthermore, the YM gauge condition
corresponding to the world-sheet Feynman-Siegel gauge is now known:
The usual gauge-fixing function $\partial \cdot A$ of the Fermi-Feynman gauge is modified to
\be \partial \cdot A + \frac{i}{8} [ \bo A_{\mu} +
\partial_{\mu} (\partial \cdot A) , A^{\mu}] + \frac{1}{4} \{ \partial \cdot A, \phi \}
+\frac{i}{8} \{ \widetilde{C}, C \} - \frac{i}{16}  \{ \bo \widetilde{C}, C \} -
\frac{i}{16}  \{ \partial_{\mu} \widetilde{C}, \partial^{\mu} C \}
\ee
The additional gauge fixing terms
simplify the $F^3$ and $\phi F^2$ interactions, and make the gauge fixed action symmetric in ghosts and antighosts \cite{driebach}.

\section{High orders of Regge slope}\label{5}

This is not the end of the story, because we only made the  action manifestly gauge
invariant to lowest order in the Regge slope expanded from the
nonlocal factors. Remember, in the 3-string vertex in (\ref{V3witten}),
the Neumann coefficients $\frac{1}{2} N_{00}^{rr}=-\lambda$ will contribute nonlocal factors to interactions.
That means
the full interaction will have the form of replacing each (anti)field $\psi_i$ in (\ref{S_3}) by
$e^{-\lambda (p_i^2 + m_i^2)} \psi_i$. But the above canonical transformations can
be performed in the same way except that the (anti)fields
$\psi_i$ in $G_g$ and $G_0$ are replaced by $e^{\lambda (\bo_i - m_i^2)} \psi_i$. Then we will get the
full action as in (\ref{S_gi}) while attaching the factor $e^{\lambda (\bo_i - m_i^2)}$ to each (anti)field
$\psi_i$ in the interaction part:
\be S^{full}_2 = -\frac{1}{2} \phi (\bo +2) \phi + \frac{1}{4} \partial_{[\mu} A_{\nu]}
\partial^{[\mu} A^{\nu]} - 2i (\partial_{\mu} C) A^{*\mu}
-2 (\widetilde{C}^*)^2 \ee
\bea
S^{full}_3 (\lambda) &=& \frac{i}{2} \partial_{\mu} \hat{\phi} [\hat{A}^{\mu} , \hat{\phi}] +
\frac{i}{4} \widehat{F}_{\mu \nu} [\hat{A}^{\mu} , \hat{A}^{\nu} ] + \frac{1}{3} \hat{\phi}^3 +
[\hat{A}_{\mu} , \hat{C} ] \hat{A}^{*\mu} + \{ \hat{C}, \hat{\phi}^* \} \hat{\phi}
+ \hat{C}^2 \hat{C}^* \nonumber \\
&& + 2 \hat{\phi} \hat{F}_{\mu \nu} \hat{F}^{\mu \nu}
- \frac{4}{3} \hat{F}_{\mu}^{\nu} \hat{F}_{\nu}^{\lambda} \hat{F}_{\lambda}^{\mu}
\eea
where
\be \hat{F}_{\mu \nu} = \partial_{[\mu} \hat{A}_{\nu]} \ee
and
\bea & \hat{\phi} = e^{\lambda (\bo + 2)} \phi ,
\quad \hat{\phi}^* = e^{\lambda (\bo + 2)} \phi^*, \quad
\hat{A} = e^{\lambda \bo} A , \quad \hat{A}^* = e^{\lambda \bo} A^* \nonumber \\
& \hat{C} = e^{\lambda \bo} C , \quad \hat{C}^* = e^{\lambda \bo} C^*, \quad
\hat{\widetilde{C}} = e^{\lambda \bo} \widetilde{C} , \quad
\hat{\widetilde{C}}^* = e^{\lambda \bo} \widetilde{C}^*
\eea

We now perform more field redefinitions to push these
nonlocal factors into higher-point interactions and restore the usual gauge invariant
action up to 3-point terms.
Let's first expand the exponential factor $e^{\lambda (\bo_i - m_i^2)}$ to the first order. Then
there are extra terms like $\phi^2 [ \lambda (\bo + 2) \phi]$ from $\frac{1}{3} \phi^3$ to
be absorbed. The naive guess is making the field redefinition through $G = \lambda \phi^2 \phi^*$,
which will give a counter term through the antibracket:
\be \delta S_3 = (G, S_2) = ( \lambda \phi^2 \phi^* , -\frac{1}{2} \phi (\bo + 2) \phi )
= - \lambda \phi^2 [(\bo + 2) \phi] \ee
where we use $S_2$ to represent the free part and $S_3$ the interaction part in (\ref{S_gi})
(to lowest order in Regge slope).

Fortunately, it turns out this is almost the right guess. To the first order in Regge slope,
the redefinition should come through
\be\label{wholeG} G = \lambda (\phi^* , S_3) \phi^* + \lambda (A^*_{\mu} , S_3) A^*_{\mu}
+ \lambda (C^*, S_3) (-\frac{i}{2}) (\partial \cdot A)
+ \lambda (\partial \cdot A , S_3) (\frac{i}{2} C^*)
\ee
Then
\bea\label{1order} (\lambda (\phi^* , S_3) \phi^* , S_2) &=& \lambda (\phi^* , S_3) (\phi^* , S_2) +
\lambda \phi^* ( (\phi^*, S_3), S_2) \nonumber \\ &=& -\lambda (\phi^*, S_3) (\bo + 2) \phi +
\lambda \phi^* (S_2 , (\phi^* , S_3))
\eea
Using the properties of antibrackets in (\ref{property}) and the gauge invariant condition
$(S_3 , S_2)$ $= 0$,
\bea && -(S_2 , (\phi^* , S_3)) + (\phi^* , (S_3 , S_2)) + (S_3 , (S_2, \phi^*)) = 0 \nonumber \\
& \Rightarrow & (S_2 , (\phi^* , S_3)) = (S_3 , (S_2, \phi^*)) = (S_3 , (\bo + 2) \phi) \nonumber \\
& & = (-[C , \phi] \phi^* , (\bo + 2) \phi) = - [C , \phi] (\bo +2 )
\eea
Thus (\ref{1order}) gives
\bea (\lambda (\phi^* , S_3) \phi^* , S_2) = -\lambda (\phi^*, S_3) (\bo + 2) \phi -
\lambda \{ C , (\bo +2) \phi^* \} \phi
\eea
which will cancel the additional terms from the first-order expansions of
$e^{\lambda (\bo - m^2)}$ for $\phi$'s and $\phi^*$'s in the 3-point
interactions. Similar calculations show that $G$ in (\ref{wholeG}) does cancel
all additional terms from the first-order expansions of
$e^{\lambda (\bo_i - m_i^2)}$ for all (anti)fields:
$\phi, \phi^*, A_{\mu} , A^*_{\mu}, C, C^*, \widetilde{C}, \widetilde{C}^* $ in $S^{full}_3$.

Basically, we can do it order by order, and here is the field redefinition for all orders:
\bea G &=& (\phi^*, \int_0^{\lambda} d \alpha S^{full}_3 (\alpha) ) \phi^* +
(A^*_{\mu}, \int_0^{\lambda} d \alpha S^{full}_3 (\alpha)) (A^*)^{\mu} \nonumber \\ &&
+ (C^*, \int_0^{\lambda} d \alpha S^{full}_3 (\alpha)) (-\frac{i}{2})
(\partial \cdot A) + (\partial \cdot A , \int_0^{\lambda}
d \alpha S^{full}_3 (\alpha)) (\frac{i}{2} C^*)
\eea
The integral is easy to perform:
\be \int_0^{\lambda} d \alpha S^{full}_3 (\alpha) = \frac{1}{(\bo_1 - m_1^2)
+ (\bo_2 - m_2^2) + (\bo_3 - m_3^2)}
(e^{\lambda (\bo_1 - m_1^2)+ \lambda (\bo_2 - m_2^2) + \lambda (\bo_3 - m_3^2)} -1) S_3 \ee
where the indices $1,2,3$ indicate the three fields in each term of $S_3$.
The proof is very similar to the first-order case and
we won't bother to give the details here.

Then we will have N-point interactions for
any big N just from a 3-string interaction in SFT. This is because in
the above calculation we only accounted for corrections up
to 3-point, while the full transformed action should be
\be e^{{\mathcal L_G}} S = S + (G, S) + \frac{1}{2!} (G, (G, S)) + \cdots \ee
Essentially, we can perform this mechanism perturbatively in
higher-point interactions. We have not studied whether the nonlocal interactions can be eliminated at any finite order of perturbation, or whether this procedure is consistent nonperturbatively.

\section{Discussion and conclusions}\hypertarget{end}{}

In this paper, we computed the ZJBV  action for Witten's open string field
theory for tachyons and massless vectors, including all ghosts and antifields. We find after some canonical transformations that the
action up to 3-point terms is just the usual Yang-Mills one plus $\phi F^2$ and $F^3$ interactions
as in (\ref{S_gi}), which is explicitly gauge invariant now. The gauge
condition in field theory which corresponds to the Feynman-Siegel gauge on the
world-sheet is also known. Furthermore, there are no nonlocal interactions in the action
up to 3-point terms. (A higher-point analysis would require analyzing the massive fields, since redefinitions of massive fields appearing in propagators, in 4-point and higher diagrams, will produce new local terms for massless fields on external lines.)  We pushed these nonlocal factors in 3-point interactions to higher-point
interactions.  It may be possible that all such explicit factors can be eliminated in the complete action, so that all ``nonlocality" can be attributed to the presence of higher-spin fields.

\section*{Acknowledgement}

It is a great pleasure to thank Leonardo Rastelli for his helpful discussion.
This work is supported in part by National Science Foundation
Grant No.\ PHY-0354776.


\begin{thebibliography}{99}

\bibitem{fs1}
H. Feng and W. Siegel, \hhref{0310070}, {\it Nucl. Phys.} {\bf B683} (2004) 168.

\bibitem{fs2}
H. Feng and W. Siegel, \hhref{0409187}, {\it Phys. Rev.} {\bf D71} (2005) 106001.

\bibitem{Goddard}
P. Goddard, J. Goldstone, C. Rebbi, and C. Thorn, {\it Nucl. Phys.} {\bf B56} (1973) 109;\\
S. Mandelstam, {\it Nucl. Phys.} {\bf B64} (1973) 205, {\bf B83} (1974) 413,
{\it Phys. Rep.} {\bf 13} (1974) 259;\\
M. Kaku and K. Kikkawa, {\it Phys. Rev.} {\bf D10} (1974) 1110, 1823.

\bibitem{Green}
M.B. Green and J.H. Schwarz, {\it Nucl. Phys.} {\bf B218} (1983) 43;\\
M.B. Green, J.H. Schwarz, and L. Brink, {\it Nucl. Phys.} {\bf B219} (1983) 437;\\
M.B. Green and J.H. Schwarz, {\it Nucl. Phys.} {\bf B243} (1984) 475.

\bibitem{witten}
E. Witten, {\it Nucl. Phys.} {\bf B268} (1986) 253.

\bibitem{Siegel}
W. Siegel, {\it Phys. Lett.} {\bf B142} (1984) 276; {\bf B151} (1985) 391;\\
W. Siegel and B. Zwiebach, {\it Nucl.Phys.} {\bf B263} (1986) 105; \\
T. Banks and M.E. Peskin, {\it Nucl.Phys.} {\bf B264} (1986) 513.

\bibitem{Gross-Jevicki-123}
D.\ J.\ Gross and A.\ Jevicki, \NP {\bf B283} (1987) 1, {\bf B287}
(1987) 225, {\bf B293} (1987) 29;\\
E. Cremmer, C.B. Thorn, and A. Schwimmer, {\it Phys. Lett.} {\bf 179B} (1986) 57.

\bibitem{lpp}
A.\ Leclair, M.\ E.\ Peskin, and C.\ R.\ Preitschopf,
{\it Nucl. Phys.} {\bf B317} (1989) 411, 464.

\bibitem{Gaberdiel-Zwiebach}
M.\ R.\ Gaberdiel and B.\ Zwiebach, \hhref{9705038},
\NP {\bf B505} (1997) 569; \hhref{9707051},
\PL {\bf B410} (1997) 151.

\bibitem{ZJ}
J. Zinn-Justin, {\it in Trends in elementary particle theory}, eds.
H. Rollnik and K. Dietz ({\bf Springer-Verlag}, 1975) p. 2.

\bibitem{BV}
I.A. Batalin and G.A. Vilkovisky, {\it Phys. Lett.} {\bf 102B} (1983) 27, {\bf 120B} (1983) 166,
{\it Phys. Rev.} {\bf D28} (1983) 2567, {\bf D30} (1984) 508, {\it Nucl. Phys.} {\bf B234} (1984) 106,
{\it J. Math. Phys.} {\bf 26} (1985) 172:

\bibitem{Siegel1}
W. Siegel and B. Zwiebach, {\it Nucl. Phys.} {\bf B299} (1988) 206.

\bibitem{Taylor}
W.\ Taylor and B.\ Zwiebach, \hhref{0311017},
{\it Boulder 2001, Strings, branes and extra dimensions} (2003) 641.

\bibitem{driebach}
  B.~Zwiebach,
  \hhref{0010190}.

\end{thebibliography}
\end{document}